\documentclass{PoS}

\def\Sl#1{\rlap{\hbox{$\mskip 3 mu /$}}#1}      
\def\tbeta{\tilde\beta}
\def\Eq#1{Eq.~(\ref{#1})}

\title{Beta function of three-dimensional QED}

\ShortTitle{Three-dimensional QED}

\author{\speaker{Benjamin Svetitsky}, Ohad Raviv, and Yigal Shamir\\
        Raymond and Beverly School of Physics and Astronomy,
        Tel Aviv University, 69978 Tel Aviv, Israel\\
        E-mail: \email{bqs@julian.tau.ac.il}}

\abstract{We have carried out a Schr\"odinger-functional calculation for the Abelian gauge theory with $N_f=2$ four-component fermions in three dimensions. We find no fixed point in the beta function, meaning that the theory is confining rather than conformal.}

\FullConference{The 32nd International Symposium on Lattice Field Theory,\\
		23-28 June, 2014\\
		Columbia University, New York, NY}

\begin{document}
\section{Introduction}
It's been some time since three-dimensional QED, or QED3, has appeared at a Lattice meeting~\cite{Strouthos:2008kc}.
Initial interest in the theory came from its connection to finite-temperature QCD via dimensional reduction~\cite{Pisarski:1984dj}.
It has since acquired a number of connections to condensed-matter systems such as the quantum Hall effect \cite{Ludwig} and high-$T_c$ superconductors~\cite{Franz}.
Our path to the theory came from the fact that it presents similar issues to those of borderline-conformal gauge theories in four dimensions~\cite{Kuti,YAoki}.
Thus we have approached it \cite{Raviv,Raviv:2014xna} with the machinery that we have applied to non-Abelian theories with fermions in assorted representations of the gauge group~\cite{Svetitsky:2013px,DeGrand:2013uha,DeGrand:2013yja}.

For definiteness, here is the theory's action:
\begin{equation}
S=\int d^3x\,\Biggl(\frac1{4e_0^2}F_{\mu\nu}F^{\mu\nu}
+\sum_{i=1}^{N_f}\bar\psi_i \Sl D\psi_i\Biggr),
\end{equation}
where $\psi$ is a massless four-component Dirac field, replicated $N_f$ times.
The question we confront is whether the IR physics of the theory is that of confinement or of conformality.
What makes this theory a difficult one to study is that in three dimensions one faces severe infrared problems, leading to sensitivity to the volume that makes interpretation of lattice results \cite{Hands:2002dv,Hands:2004bh} less than straightforward~\cite{Gusynin:2003ww,Fischer:2004nq}.

There is nothing non-Abelian here.
All we have is charged fermions, a two-dimensional ($\log r$) Coulomb potential, and a transverse photon.
The complication comes from screening by the massless charges.
Does the confining potential win, or do the charges screen it?
Previous work \cite{Appelquist:2004ib} shows that there are two plausible regimes:
\begin{enumerate}
\item
For small $N_f$, there is confinement and mass generation for the charges, with $m\sim e^2$.
\item
For large $N_f$, screening wins.
\end{enumerate}

To explain further, let us focus on the running coupling $e^2(q)$.
Since the one-loop diagram involves screening, just like QED4, we have the perturbative form
\begin{equation}
\frac{de^2}{d\log q}=N_fb_1 e^4/q+\cdots,
\end{equation}
with $b_1>0$.
If we define a dimensionless coupling $g^2(q)=e^2/q$, this becomes
\begin{equation}
\frac{dg^2}{d\log q}=-g^2+N_fb_1 g^4+\cdots.
\label{beta}
\end{equation}
Shades of QCD!
The first term, typical of a super-renormalizable theory, drives the theory towards strong coupling in the IR, inviting a condensate $\langle\bar\psi\psi\rangle$ and a dynamical mass for the fermions, which therefore decouple at long distances and leave us with a logarithmic, confining potential.
If $N_f$ is large, though, the coupling only runs as far as a fixed point at $g^2=(N_fb_1)^{-1}$.
At long distance we see conformal physics, with no length scale (and no particles).

If small-$N_f$ physics differs from large $N_f$, there must be a critical value $N_{\rm cr}$ in between.
Analytical calculations have converged \cite{Fischer:2004nq} to a value in the neighborhood of $N_{\rm cr}=4$.
Upper bounds on $N_{\rm cr}$, rather larger than this, have been derived from the $F$-theorem governing monotonicity in renormalization group flows~\cite{Grover:2012sp}. 
I will present our study of $N_f=2$, which falls into line with these results.%
\footnote{Recently the possibility has been raised \cite{Braun:2014wja} of a region in $N_f$ intermediate between mass generation at small $N_f$ and conformality at large $N_f$.
I have nothing to say about this, except that it's interesting.}

\section{Calculating the $\beta$ function}

The Schr\"odinger functional method \cite{Luscher:1992an,Luscher:1993gh} has been widely used to define a running coupling for QCD and QCD-like theories in four dimensions.
Its outstanding feature is that it uses the finite volume of the system to define the scale at which the coupling runs.
Thus in QED3, plagued by infrared difficulties, the finite volume of a lattice calculation is turned from a hindrance into a tool.

We define our theory in a three-dimensional Euclidean box of dimension $L$.
We fix simple boundary conditions on the gauge field at $t=0$ and $L$, namely $A_x=A_y=\pm\phi/L$.
This amounts to imposing a uniform background field $E_x=E_y=-2\phi/L^2$.
Note that $L$ is the only scale, so that the eventual running coupling will be $g^2(L)$.
The latter is derived from a calculation of the free energy $\Gamma=-\log Z$ in the presence of the background field.
Comparison to the classical action gives the effective coupling via
\begin{equation}
\Gamma = \frac1{e^2(L)}\int d^3x\,\frac14 F_{\mu\nu}F^{\mu\nu}.
\label{comparison}
\end{equation}
Since the integral in \Eq{comparison} is just $\frac12L^3E^2=8\phi^2/L$, a calculation of $\Gamma$ gives directly%
\footnote{More precisely, one calculates the derivative $d\Gamma/d\phi$, which is some Green function of the theory.}
the running coupling $g^2(L)=e^2(L)L$ and hence the beta function.

A one-loop calculation shows what we might look for.
From \Eq{beta} we define the beta function for $u\equiv1/g^2$,
\begin{equation}
\tbeta(u)\equiv\frac{d(1/g^2)}{d\log L}=-\frac1{g^2}+N_fb_1 +O(g^2),
\end{equation}
a straight line that crosses zero at $u=N_fb_1$---the one-loop fixed point.

\section{Lattice calculation}

\begin{figure}
\begin{center}
\includegraphics[width=.6\columnwidth,clip]{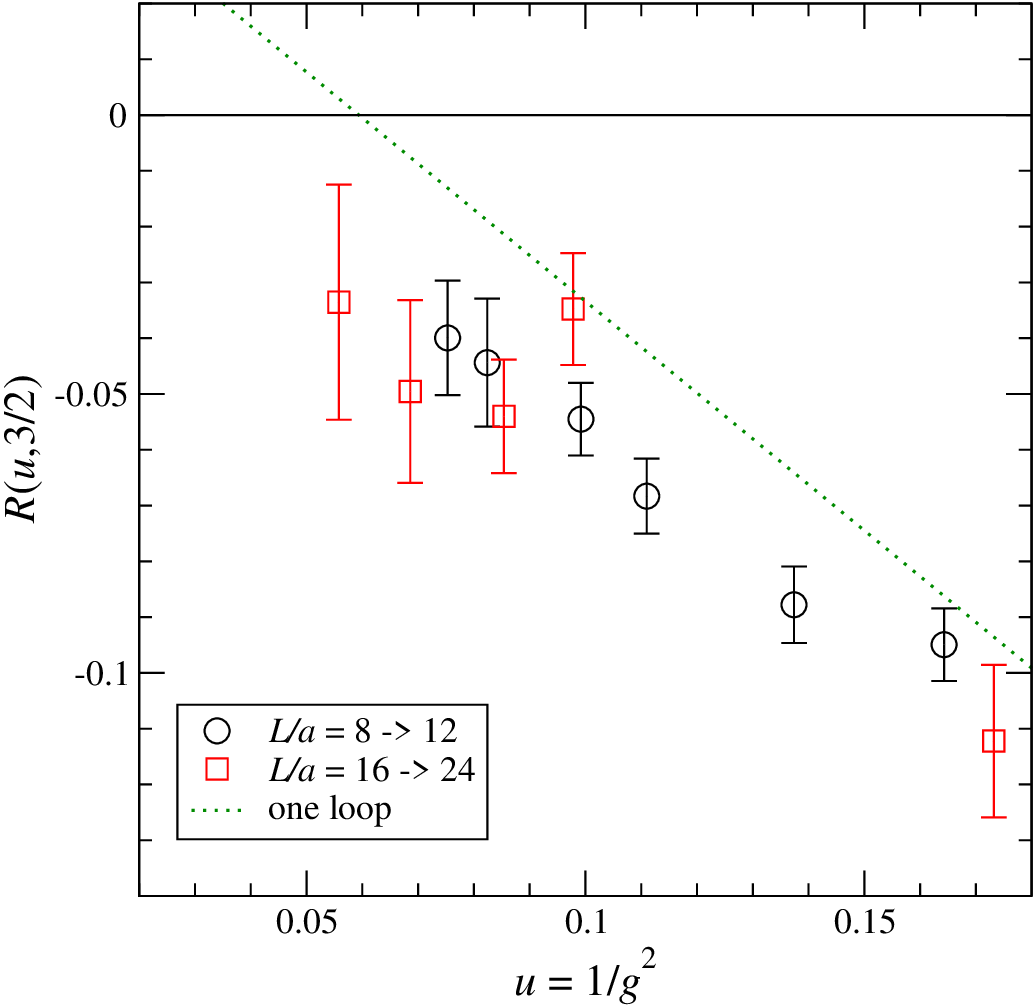}
\end{center}
\caption{The discrete beta function $R(u,s)$ for scale factor $s=3/2$.
\label{RDBF}}
\end{figure}

We use a non-compact gauge field (no instantons!) with Wilson--clover fermions and nHYP smearing,
\begin{equation}
S=\frac{\beta}2\sum_{\scriptstyle n\atop\scriptstyle \mu<\nu}(\nabla\times A)_{n\mu\nu}^2+\bar\psi D\psi,
\end{equation}
with bare coupling $\beta=1/(e_0^2a)$, on a lattice of dimension $L=Na$.
We fix $\kappa=\kappa_c(\beta)$ to enforce masslessness.
The simulation, as described above, gives $u(L)$ directly.
We can compare calculations on two lattices of size $L$ and $sL$, keeping $(\beta,\kappa)$ fixed in order to keep $a$ fixed.
This gives the ``rescaled'' discrete beta function,
\begin{equation}
R(u,s)\equiv\frac{u(sL)-u(L)}{\log s},
\end{equation}
shown in Fig.~\ref{RDBF}.
($R$ tends to the beta function $\tbeta$ as $s\to1$.)
Two sets of data are shown in the figure, for two different lattice sizes.
Remember that $u=1/g^2$ is the running coupling at the physical scale $L$.
This is renormalization: Fixing $u$ means fixing $L$.
Increasing $L/a$ at fixed $u$ means that $L$ is fixed while $a$ is decreased.
Thus the two sets of data points represent two different lattice spacings.
Fig.~\ref{RDBF} is a first look at the beta function, which apparently avoids the perturbative fixed point and levels off in strong coupling.

\section{Continuum extrapolation}

\begin{figure}
\begin{center}
\includegraphics[width=.6\columnwidth,clip]{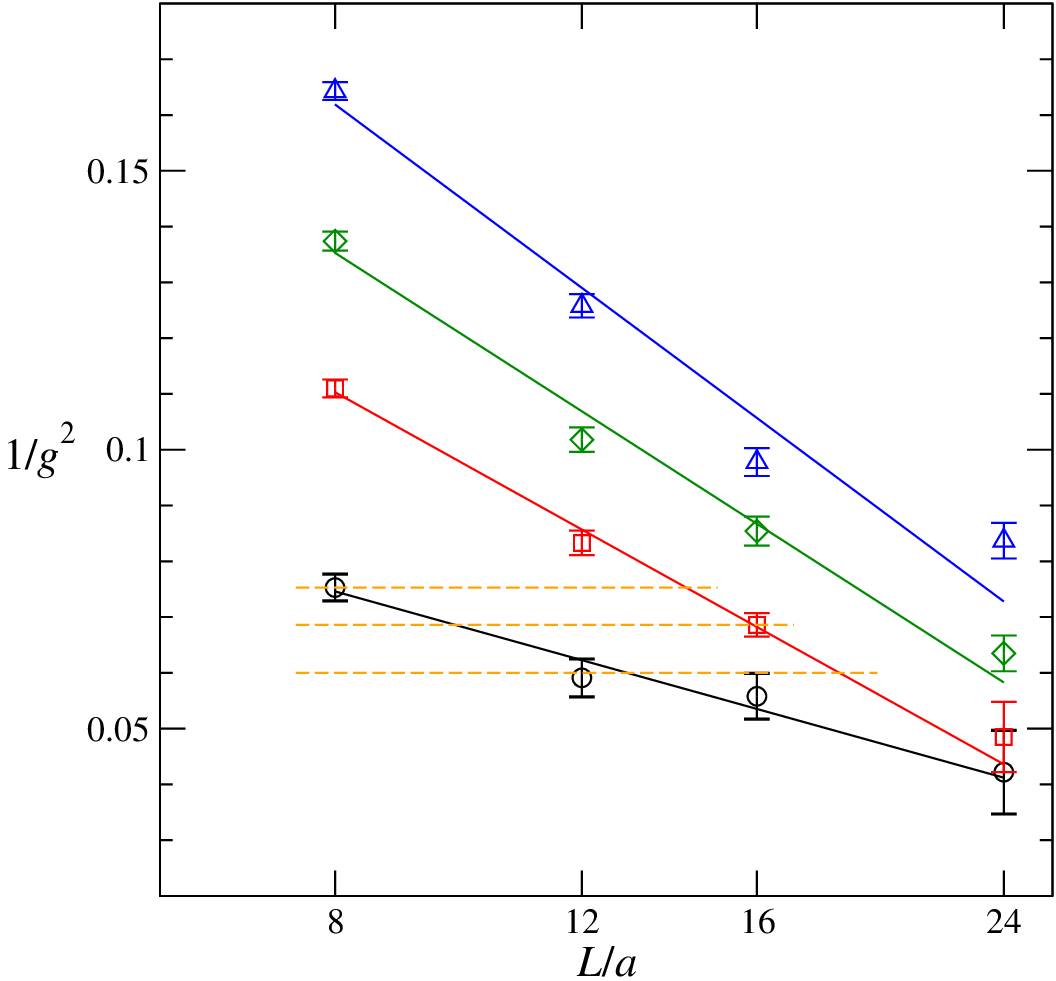}\\
\end{center}
\caption{Running coupling vs.~lattice size at fixed bare coupling $\beta$.
Top to bottom: $\beta= 1.0$, 0.8, 0.6,~0.4.
\label{a3dcouplings}}
\end{figure}

For more systematic analysis  of the dependence on lattice spacing, we carry out an analysis that is close in spirit to that used in most Schr\"odinger functional calculations.
We plot in Fig.~\ref{a3dcouplings} the coupling $1/g^2$ against $\log L$ for fixed bare coupling $\beta$.
We are looking for a leveling off in the beta function at strong coupling.
If the beta function is constant, the coupling will change by a fixed amount for each change in $\log L$ at fixed lattice spacing.
Then each group of data points, at fixed $\beta$, will lie on a straight line whose slope is the beta function.
We see that this works (approximately) only for the two strongest bare couplings, that is, for the bottom two groups of data points.

\begin{figure}
\begin{center}
\includegraphics[width=.6\columnwidth,clip]{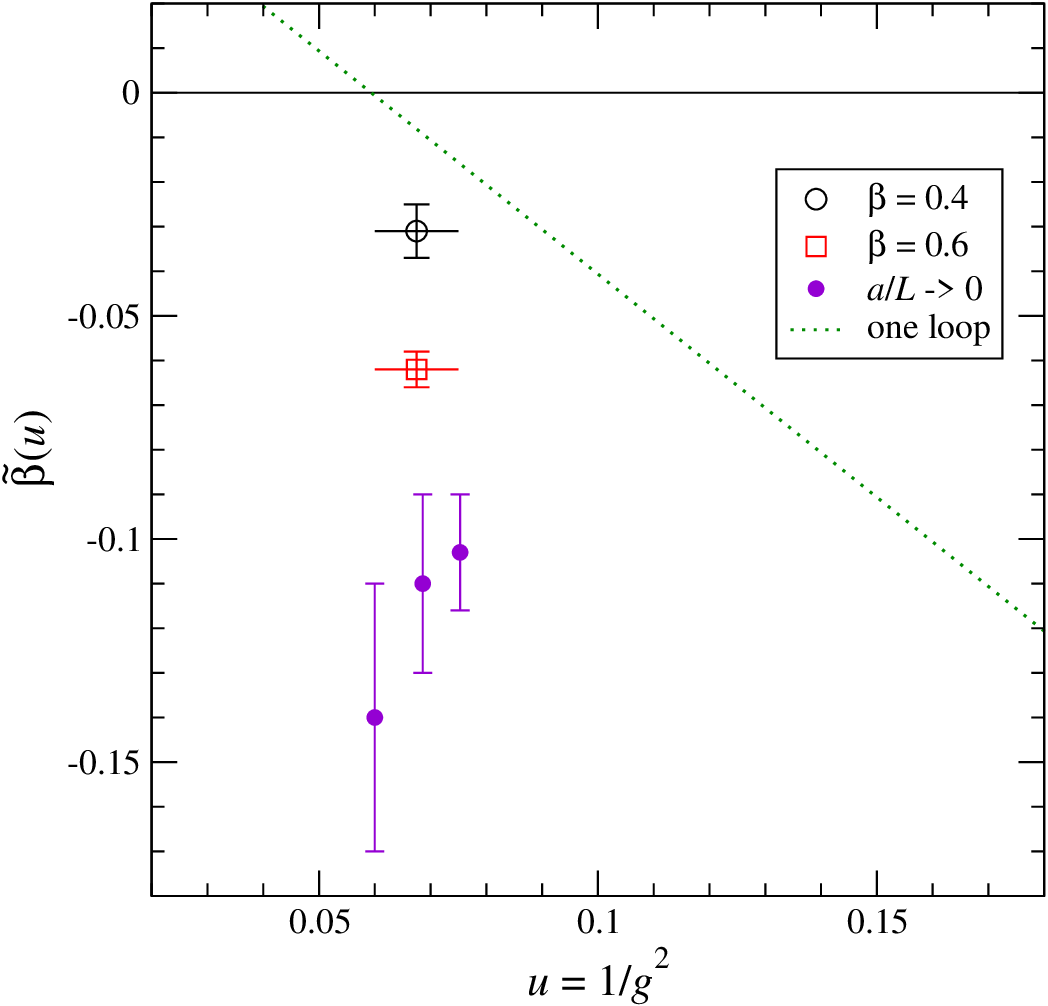}\\
\end{center}
\caption{Extrapolation of the beta function to the continuum for the three couplings marked by horizontal lines in Fig.~2.
\label{slope_method}}
\end{figure}

The horizontal lines in Fig.~\ref{a3dcouplings} show that at fixed $g^2$, which means fixed physical size $L$, we have two different slopes at two different bare couplings $\beta$---which means two different lattice spacings $a$.
Thus we can extrapolate to $a/L=0$, giving a continuum extrapolation of the slope, that is, the beta function.
Fig.~\ref{slope_method} shows this extrapolation at the strongest couplings we can reach.
Again, the avoidance of the one-loop zero is clear.
In fact, comparison to Fig.~\ref{RDBF} shows that this behavior is enhanced by the continuum extrapolation.
The conclusion is that QED3 with $N_f=2$ confines.

Let me end with the comment that the one-loop beta function in this theory is very different from that of the near-conformal theories in four dimensions that we have studied in the past.
Correspondingly, the results of our numerical calculations differ qualitatively as well.
The slow running of the coupling in the four-dimensional theories required a rather difficult procedure of extrapolation to the continuum limit~\cite{DeGrand:2013yja}, and the Monte Carlo data available to us allowed only limited success.
The present analysis of QED3 is more straightforward.

\section*{Acknowledgments}
This work was supported in part by the Israel Science Foundation under Grants
No.~423/09 and No.~1362/08 and by the European Research
Council under Grant No.~203247.
I thank Christian Fischer and Sinya Aoki for conversations during and after the conference.

\end{document}